\documentclass[11pt]{article}
\begin{document}
%\begin{document}
%\special{papersize=8.26in,11.69in}
%\textwidth15.0cm
%\textheight22.0cm
%\baselineskip1.0cm
%\setlength{\topmargin}{-1cm}
%\addtolength{\textheight}{1cm}
%\oddsidemargin+1.2cm
%\evensidemargin-1.2cm
\pagestyle{empty}

%%%%%%%%%%%%%%% insert actual file mydefs.sty %%%%%%%%%%%%%%%%%%%%%%
%-------------------------------------------------------------------------
%% My own definitions
% From cv.sty
%
% Title page
 
\def\appendix{\par\clearpage
  \setcounter{section}{0}
  \setcounter{subsection}{0}
  \@addtoreset{equation}{section}
  \def\@sectname{Appendix~}
  \def\theequation{\thesection\arabic{equation}}
  \def\thesection{\Alph{section}}}
 
% Figures
\def\thefigures#1{\par\clearpage\section*{Figures\@mkboth
  {FIGURES}{FIGURES}}\list
  {Fig.~\arabic{enumi}.}{\labelwidth\parindent\advance
\labelwidth -\labelsep
      \leftmargin\parindent\usecounter{enumi}}}
\def\figitem#1{\item\label{#1}}
\let\endthefigures=\endlist
 
% Tables
\def\thetables#1{\par\clearpage\section*{Tables\@mkboth
  {TABLES}{TABLES}}\list
  {Table~\Roman{enumi}.}{\labelwidth-\labelsep
      \leftmargin0pt\usecounter{enumi}}}
\def\tableitem#1{\item\label{#1}}
\let\endthetables=\endlist
 
% Put period after section number and allow for APPENDIX prefix.
\def\@sect#1#2#3#4#5#6[#7]#8{\ifnum #2>\c@secnumdepth
     \def\@svsec{}\else
     \refstepcounter{#1}\edef\@svsec{\@sectname\csname the#1\endcsname
.\hskip 1em }\fi
     \@tempskipa #5\relax
      \ifdim \@tempskipa>\z@
        \begingroup #6\relax
          \@hangfrom{\hskip #3\relax\@svsec}{\interlinepenalty \@M #8\par}
        \endgroup
       \csname #1mark\endcsname{#7}\addcontentsline
         {toc}{#1}{\ifnum #2>\c@secnumdepth \else
                      \protect\numberline{\csname the#1\endcsname}\fi
                    #7}\else
        \def\@svse=chd{#6\hskip #3\@svsec #8\csname #1mark\endcsname
                      {#7}\addcontentsline
                           {toc}{#1}{\ifnum #2>\c@secnumdepth \else
                             \protect\numberline{\csname the#1\endcsname}\fi
                       #7}}\fi
     \@xsect{#5}}
 
\def\@sectname{}
%
%                 M A T E X
%
%       This defines et al., i.e., e.g., cf., etc.
\def\eg{\hbox{\it e.g.}}        \def\cf{\hbox{\it cf.}}
\def\etal{\hbox{\it et al.}}
\def\dash{\hbox{---}}
%       common physics symbols
\def\bR{\mathop{\bf R}}
\def\bC{\mathop{\bf C}}
\def\eq#1{{eq. \ref{#1}}}
\def\eqs#1#2{{eqs. \ref{#1}--\ref{#2}}}
\def\lie{\hbox{\it \$}} % fancy L for the Lie derivative
\def\partder#1#2{{\partial #1\over\partial #2}}
\def\secder#1#2#3{{\partial^2 #1\over\partial #2 \partial #3}}
\def\abs#1{\left| #1\right|}
\def\ltap{\ \raisebox{-.4ex}{\rlap{$\sim$}} \raisebox{.4ex}{$<$}\ }
\def\gtap{\ \raisebox{-.4ex}{\rlap{$\sim$}} \raisebox{.4ex}{$>$}\ }
% \contract is a differential geometry contraction sign _|
\def\contract{\makebox[1.2em][c]{
        \mbox{\rule{.6em}{.01truein}\rule{.01truein}{.6em}}}}
% double-headed superior arrow added 9.2.86
%
% commutator added 11.14.86
\def\com#1#2{
        \left[#1, #2\right]}
%
%these written by orlando alvarez
% ************************************************************
%       The following macros were written by Chris Quigg.
%       They create bent arrows and can be used to write
%       decays such as pi --> mu + nu
%                              --> e nu nubar
%
\def\bentarrow{\:\raisebox{1.3ex}{\rlap{$\vert$}}\!\rightarrow}
\def\longbent{\:\raisebox{3.5ex}{\rlap{$\vert$}}\raisebox{1.3ex}%
        {\rlap{$\vert$}}\!\rightarrow}
\def\onedk#1#2{
        \begin{equation}
        \begin{array}{l}
         #1 \\
         \bentarrow #2
        \end{array}
        \end{equation}
                }
\def\dk#1#2#3{
        \begin{equation}
        \begin{array}{r c l}
        #1 & \rightarrow & #2 \\
         & & \bentarrow #3
        \end{array}
        \end{equation}
                }
\def\dkp#1#2#3#4{
        \begin{equation}
        \begin{array}{r c l}
        #1 & \rightarrow & #2#3 \\
         & & \phantom{\; #2}\bentarrow #4
        \end{array}
        \end{equation}
                }
\def\bothdk#1#2#3#4#5{
        \begin{equation}
        \begin{array}{r c l}
        #1 & \rightarrow & #2#3 \\
         & & \:\raisebox{1.3ex}{\rlap{$\vert$}}\raisebox{-0.5ex}{$\vert$}%
        \phantom{#2}\!\bentarrow #4 \\
         & & \bentarrow #5
        \end{array}
        \end{equation}
                }
%
%
%abbreviated journal names
%
\def\ap#1#2#3{     {\it Ann. Phys. (NY) }{\bf #1} (19#2) #3}
\def\arnps#1#2#3{  {\it Ann. Rev. Nucl. Part. Sci. }{\bf #1} (19#2) #3}
\def\npb#1#2#3{    {\it Nucl. Phys. }{\bf B #1} (19#2) #3}
\def\plb#1#2#3{    {\it Phys. Lett. }{\bf B #1} (19#2) #3}
\def\prd#1#2#3{    {\it Phys. Rev. }{\bf D #1} (19#2) #3}
\def\prep#1#2#3{   {\it Phys. Rep. }{\bf #1} (19#2) #3}
\def\prl#1#2#3{    {\it Phys. Rev. Lett. }{\bf #1} (19#2) #3}
\def\ptp#1#2#3{    {\it Prog. Theor. Phys. }{\bf #1} (19#2) #3}
\def\rmp#1#2#3{    {\it Rev. Mod. Phys. }{\bf #1} (19#2) #3}
\def\zpc#1#2#3{    {\it Zeitschr. f{\"u}r Physik }{\bf C #1} (19#2) #3}
\def\mpla#1#2#3{   {\it Mod. Phys. Lett. }{\bf A #1} (19#2) #3}
\def\sjnp#1#2#3{   {\it Sov. J. Nucl. Phys. }{\bf #1} (19#2) #3}
\def\yf#1#2#3{     {\it Yad. Fiz. }{\bf #1} (19#2) #3}
\def\jetpl#1#2#3{  {\it JETP Lett. }{\bf #1} (19#2) #3}
\def\ib#1#2#3{     {\it ibid. }{\bf #1} (19#2) #3}
\def\lmp#1#2#3{    {\it Lett. Math. Phys. }{\bf #1} (19#2) #3}
\def\app#1#2#3{    {\it Act. Phys. Pol.}{\bf B #1} (19#2) #3}
\def\cqg#1#2#3{    {\it Class. Quant. Grav. }{\bf #1} (19#2) #3}  
\newcommand{\nc}{\newcommand}
\nc{\spa}[3]{\left\langle#1\,#3\right\rangle}
\nc{\spb}[3]{\left[#1\,#3\right]}
\nc{\ksl}{\not{\hbox{\kern-2.3pt $k$}}}
\nc{\hf}{\textstyle{1\over2}}
\nc{\pol}{\varepsilon}
\nc{\tq}{{\tilde q}}
\nc{\esl}{\not{\hbox{\kern-2.3pt $\pol$}}}
\newcommand{\1}{{\'\i}}
\newcommand{\be}{\begin{equation}}
\newcommand{\ee}{\end{equation}\noindent}
\newcommand{\bear}{\begin{eqnarray}}
\newcommand{\ear}{\end{eqnarray}\noindent}
\newcommand{\benn}{\begin{enumerate}}
\newcommand{\enn}{\end{enumerate}}
\newcommand{\no}{\noindent}
\date{}
\renewcommand{\theequation}{\arabic{section}.\arabic{equation}}
\renewcommand{\arraystretch}{2.5}
\newcommand{\GeV}{\mbox{GeV}}
\newcommand{\cL}{\cal L}
\newcommand{\D}{\cal D}
\newcommand{\Dhalf}{{D\over 2}}
\newcommand{\Det}{{\rm Det}}
\newcommand{\PP}{\cal P}
\newcommand{\G}{{\cal G}}
\def\R{1\!\!{\rm R}}
\def\Eins{\mathord{1\hskip -1.5pt
\vrule width .5pt height 7.75pt depth -.2pt \hskip -1.2pt
\vrule width 2.5pt height .3pt depth -.05pt \hskip 1.5pt}}
\newcommand{\symb}{\mbox{symb}}
\renewcommand{\arraystretch}{2.5}
\newcommand{\slD}{\raise.15ex\hbox{$/$}\kern-.57em\hbox{$D$}}
\newcommand{\slpartial}{\raise.15ex\hbox{$/$}\kern-.57em\hbox{$\partial$}}
\newcommand{\slG}{{{\dot G}\!\!\!\! \raise.15ex\hbox {/}}}
\newcommand{\Gd}{{\dot G}}
\newcommand{\Gund}{{\underline{\dot G}}}
\newcommand{\Gdd}{{\ddot G}}
\def\GBd12{{\dot G}_{B12}}
\def\mneg{\!\!\!\!\!\!\!\!\!\!}
\def\Mneg{\!\!\!\!\!\!\!\!\!\!\!\!\!\!\!\!\!\!\!\!}
\def\non{\nonumber}
\def\beqn*{\begin{eqnarray*}}
\def\eqn*{\end{eqnarray*}}
\def\sy{\scriptscriptstyle}
\def\footstrut{\baselineskip 12pt}
\def\square{\kern1pt\vbox{\hrule height 1.2pt\hbox{\vrule width 1.2pt
   \hskip 3pt\vbox{\vskip 6pt}\hskip 3pt\vrule width 0.6pt}
   \hrule height 0.6pt}\kern1pt}
\def\np{n_{+}}
\def\nm{n_{-}}
\def\Np{N_{+}}
\def\Nm{N_{-}}
\def\exmn{\Bigl(\mu \leftrightarrow \nu \Bigr)}
\def\slash#1{#1\!\!\!\raise.15ex\hbox {/}}
\def\dint#1{\int\!\!\!\!\!\int\limits_{\!\!#1}}
\def\bra#1{\langle #1 |}
\def\ket#1{| #1 \rangle}
\def\vev#1{\langle #1 \rangle}
\def\rightvac{\mid 0\rangle}
\def\leftvac{\langle 0\mid}
\def\dps{\displaystyle}
\def\sy{\scriptscriptstyle}
\def\half{{1\over 2}}
\def\third{{1\over3}}
\def\fourth{{1\over4}}
\def\fifth{{1\over5}}
\def\sixth{{1\over6}}
\def\seventh{{1\over7}}
\def\eigth{{1\over8}}
\def\ninth{{1\over9}}
\def\tenth{{1\over10}}
\def\pa{\partial}
\def\ddtau{{d\over d\tau}}
\def\ge{\hbox{\textfont1=\tame $\gamma_1$}}
\def\gz{\hbox{\textfont1=\tame $\gamma_2$}}
\def\gd{\hbox{\textfont1=\tame $\gamma_3$}}
\def\go{\hbox{\textfont1=\tamt $\gamma_1$}}
\def\gt{\hbox{\textfont1=\tamt $\gamma_2$}}
\def\gth{\hbox{\textfont1=\tamt $\gamma_3$}} 
\def\gf{\hbox{$\gamma_5\;$}}
\def\ie{\hbox{$\textstyle{\int_1}$}}
\def\iz{\hbox{$\textstyle{\int_2}$}}
\def\id{\hbox{$\textstyle{\int_3}$}}
\def\ldop{\hbox{$\lbrace\mskip -4.5mu\mid$}}
\def\rdop{\hbox{$\mid\mskip -4.3mu\rbrace$}}
\def\eps{\epsilon}
\def\epshalf{{\epsilon\over 2}}
\def\e{\mbox{e}}
\def\mn{{\mu\nu}}
\def\exmn{{(\mu\leftrightarrow\nu )}}
\def\ab{{\alpha\beta}}
\def\exab{{(\alpha\leftrightarrow\beta )}}
\def\g{\mbox{g}}
\def\kinb{{1\over 4}\dot x^2}
\def\kinf{{1\over 2}\psi\dot\psi}
\def\expk{{\rm exp}\biggl[\,\sum_{i<j=1}^4 G_{Bij}k_i\cdot k_j\biggr]}
\def\expp{{\rm exp}\biggl[\,\sum_{i<j=1}^4 G_{Bij}p_i\cdot p_j\biggr]}
\def\expshort{{\e}^{\half G_{Bij}k_i\cdot k_j}}
\def\expabb{{\e}^{(\cdot )}}
\def\epseps#1#2{\varepsilon_{#1}\cdot \varepsilon_{#2}}
\def\epsk#1#2{\varepsilon_{#1}\cdot k_{#2}}
\def\kk#1#2{k_{#1}\cdot k_{#2}}
\def\G#1#2{G_{B#1#2}}
\def\Gp#1#2{{\dot G_{B#1#2}}}
\def\GF#1#2{G_{F#1#2}}
\def\Dab{{(x_a-x_b)}}
\def\Dsq{{({(x_a-x_b)}^2)}}
\def\lag{( -\partial^2 + V)}
\def\PITD{{(4\pi T)}^{-{D\over 2}}}
\def\4piTD{{(4\pi T)}^{-{D\over 2}}}
\def\4piT4{{(4\pi T)}^{-2}}
\def\TintmD{{\dps\int_{0}^{\infty}}{dT\over T}\,e^{-m^2T}
    {(4\pi T)}^{-{D\over 2}}}
\def\Tintm4{{\dps\int_{0}^{\infty}}{dT\over T}\,e^{-m^2T}
    {(4\pi T)}^{-2}}
\def\Tintm{{\dps\int_{0}^{\infty}}{dT\over T}\,e^{-m^2T}}
\def\Tint{{\dps\int_{0}^{\infty}}{dT\over T}}
\def\pint{{\dps\int}{dp_i\over {(2\pi)}^d}}
\def\Dx{\dps\int{\cal D}x}
\def\Dy{\dps\int{\cal D}y}
\def\Dpsi{\dps\int{\cal D}\psi}
\def\Tr{{\rm Tr}\,}
\def\tr{{\rm tr}\,}
\def\sumij{\sum_{i<j}}
\def\freeexp{{\rm e}^{-\int_0^Td\tau {1\over 4}\dot x^2}}
\def\arraystretch{2.5}
\def\Ge{\mbox{GeV}}
\def\dA{\partial^2}
\def\DA{\sqsubset\!\!\!\!\sqsupset}
\def\FFdual{F\cdot\tilde F}
\def\mn{{\mu\nu}}
\def\rs{{\rho\sigma}}
%abbreviated journal names
%
\def\ap#1#2#3{     {\it Ann. Phys. (NY) }{\bf #1} (19#2) #3}
\def\arnps#1#2#3{  {\it Ann. Rev. Nucl. Part. Sci. }{\bf #1} (19#2) #3}
\def\npb#1#2#3{    {\it Nucl. Phys. }{\bf B #1} (19#2) #3}
\def\plb#1#2#3{    {\it Phys. Lett. }{\bf B #1} (19#2) #3}
\def\prd#1#2#3{    {\it Phys. Rev. }{\bf D #1} (19#2) #3}
\def\prep#1#2#3{   {\it Phys. Rep. }{\bf #1} (19#2) #3}
\def\prl#1#2#3{    {\it Phys. Rev. Lett. }{\bf #1} (19#2) #3}
\def\ptp#1#2#3{    {\it Prog. Theor. Phys. }{\bf #1} (19#2) #3}
\def\rmp#1#2#3{    {\it Rev. Mod. Phys. }{\bf #1} (19#2) #3}
\def\zpc#1#2#3{    {\it Zeitschr. f{\"u}r Physik }{\bf C #1} (19#2) #3}
\def\mpla#1#2#3{   {\it Mod. Phys. Lett. }{\bf A #1} (19#2) #3}
\def\sjnp#1#2#3{   {\it Sov. J. Nucl. Phys. }{\bf #1} (19#2) #3}
\def\yf#1#2#3{     {\it Yad. Fiz. }{\bf #1} (19#2) #3}
\def\nc#1#2#3{     {\it Nuovo Cim. }{\bf #1} (19#2) #3}
\def\jetpl#1#2#3{  {\it JETP Lett. }{\bf #1} (19#2) #3}
\def\ib#1#2#3{     {\it ibid. }{\bf #1} (19#2) #3}
\def\lmp#1#2#3{    {\it Lett. Math. Phys. }{\bf #1} (19#2) #3}
\def\app#1#2#3{    {\it Act. Phys. Pol.}{\bf B #1} (19#2) #3}
\def\cqg#1#2#3{    {\it Class. Quant. Grav. }{\bf #1} (19#2) #3}  
%
%\font\tame = cmmi12 scaled\magstep1
%\font\tamt = cmmi12 scaled\magstep2
%-------------------------------------------------------------------------
% To change the LaTeX pagestyle
% example  DINA4 format DESY
%\newlength{\dinwidth}
%\newlength{\dinmargin}
%\setlength{\dinwidth}{21.0cm}
%\textheight23.2cm
%\textwidth17.0cm
%\setlength{\dinmargin}{\dinwidth}
%\addtolength{\dinmargin}{-\textwidth}
%\setlength{\dinmargin}{0.5\dinmargin}
%\oddsidemargin -1.0in
%\addtolength{\oddsidemargin}{\dinmargin}
%\setlength{\evensidemargin}{\oddsidemargin}
%\setlength{\marginparwidth}{0.9\dinmargin}
%\marginparsep 8pt \marginparpush 5pt
%\topmargin -42pt
%\headheight 12pt 
%\headsep 30pt \footheight 12pt \footskip
%24pt
%-----------------------------------------------------------------------
% uncomment any of these if you want numbering to respect the sections
%
% \renewcommand{\thesection}{\arabic{section}.}
% \renewcommand{\thesubsection}{\thesection\arabic{subsection}.}
% \renewcommand{\theequation}{{\protect\thesection\arabic{equation}}}
% \renewcommand{\thetable}{{\protect{\bf \thesection\arabic{table}}}}
% \renewcommand{\thetable}{{\protect{\thesection\arabic{table}}}}
% \renewcommand{\thefigure}{{\protect\bf\thesection\arabic{figure}}}
% \renewcommand{\thefigure}{{\protect\thesection\arabic{figure}}}
% \renewcommand{\textfraction}{0}
% \renewcommand{\topfraction}{1.00}
% \renewcommand{\bottomfraction}{1.00}
% \renewcommand{\baselinestretch}{1.1}
%-----------------------------------------------------------------------
% special symbols: real numbers, unit matrix, integers
%
\def\bbbr{{\rm I\!R}}
\def\bbbone{{\mathchoice {\rm 1\mskip-4mu l} {\rm 1\mskip-4mu l}
{\rm 1\mskip-4.5mu l} {\rm 1\mskip-5mu l}}}
\def\bbbz{{\mathchoice {\hbox{$\sf\textstyle Z\kern-0.4em Z$}}
{\hbox{$\sf\textstyle Z\kern-0.4em Z$}}
{\hbox{$\sf\scriptstyle Z\kern-0.3em Z$}}
{\hbox{$\sf\scriptscriptstyle Z\kern-0.2em Z$}}}}

%%%%%%%%%%%%%%%%%%%%%%%%%%%%%%%%%%%%%%%%%%%%%%%%%%%%%%%%%%%%%%%%%%%%%%%
\renewcommand{\thefootnote}{\protect\arabic{footnote}}
%\pagestyle{plain}
%------------------------------------------------------
\hfill {\large AEI-2004-047}

\begin{center}
{\huge\bf One loop photon-graviton mixing in an electromagnetic field: Part 1}
\vskip1.3cm
{\large Fiorenzo Bastianelli}
\\[1.5ex]
{\it
Dipartimento di Fisica, Universit\`a di Bologna and INFN, Sezione di Bologna
\\
Via Irnerio 46, I-40126 Bologna, Italy
}
\vspace{.8cm}

 {\large Christian Schubert}
\\[1.5ex]
{\it 
%   \footnote{}
Max-Planck-Institut f\"ur Gravitationsphysik, Albert-Einstein-Institut,
M\"uhlenberg 1, D-14476 Potsdam, Germany
}
\\[1.5ex]
{\it
\vspace{1pt}
{\rm and}
\vspace{1pt}
\\
Department of Physics and Geology
\\
University of Texas Pan American
\\
Edinburg, TX 78541-2999, USA
}

%\centerline{\today}
\vspace{1cm}
 {\large \bf Abstract}
\end{center}
\begin{quotation}
Photon-graviton mixing in an electromagnetic field is a process of potential
interest for cosmology and astrophysics. At the tree level it has been studied
by many authors. We consider the one-loop contribution
to this amplitude involving a charged spin 0 or spin 1/2 particle in the loop and an arbitrary
constant field.  In the first part of this article, the worldline formalism is used to obtain a compact two-parameter integral representation for this amplitude, valid for arbitrary photon energies and
background field strengths. The calculation is manifestly covariant througout.
%For physical polarizations, further simplification is achieved by
%a suitable field-dependent choice of basis for the photon and graviton polarizations. 
 
\end{quotation}
\vfill\eject
\pagestyle{plain}
\setcounter{page}{1}
\setcounter{footnote}{0}

\vspace{20pt}
\section{Introduction}
\renewcommand{\theequation}{1.\arabic{equation}}
\setcounter{equation}{0}

In the presence of an external electromagnetic field, many quantum processes exist
which are forbidden in vacuum (see, e.g., \cite{raffeltbook,ditgiebook}). 
In particular, transitions between bosons of different
spin become possible \cite{gertsenshtein,zelnovbook,rafsto}.  
One such process which has been well-studied is the axion-photon mixing in a magnetic field 
\cite{rafsto,sikivie,morris}. 
Similarly, also photon-graviton mixing is possible in an external field
\cite{gertsenshtein,zelnovbook,rafsto}.
The corresponding tree level amplitude is contained in the coupling 
$h_\mn T^\mn$ of the graviton
$h_{\mu\nu}$ to the energy-momentum tensor $T^\mn$ of the electromagnetic field,

\bear
T^\mn &=& F^{\mu\alpha}F_{\,\,\alpha}^{\nu} - {1\over 4}F_{\ab}F^{\ab}\eta^\mn.
\label{defT}
\ear
Namely, when taking $F^\mn = F_{\rm ext}^\mn + f^\mn $ with $F^\mn_{\rm ext}$ the external field
strength tensor and $f^\mn $ the photon field, $h_{\mn}T^\mn $ yields the trilinear term

\bear
h_\mn\Bigl(F_{\rm ext}^{\mu\alpha}f^{\nu}_{\,\,\,\alpha} + f^{\mu}_{\,\,\alpha}\,F_{\rm ext}^{\nu\alpha}
\Bigr) - \half h^{\mu}_{\mu}F_{\rm ext}^{\ab} f_\ab.
\label{treelevel}
\ear
Due to the smallness of the gravitational coupling constant $\kappa$ this process
has attracted less attention than the axion-photon coupling. Nevertheless, its relevance
for astrophysics has been scrutinized by a number of authors (see \cite{losotr}
for a discussion of possible laboratory experiments).
In \cite{rafsto} photon-graviton
conversion near a pulsar was studied but the transition rate was found to be very small.
In \cite{magueijo,chen} it has been suggested that the same conversion due to a primordial
magnetic field could be responsible for the observed anisotropy of the cosmic microwave
background. However, \cite{cilhar} find that its effect becomes negligible for standard
cosmological magnetic fields if plasma effects are taken into account.
Renewed interest in this amplitude has been generated by the recent models
with large extra dimensions \cite{ardidv}. These models contain additional massive
Kaluza-Klein gravitons which might lead to an enhancement of the
photon-graviton conversion effect. See \cite{defuza} for a discussion of 
possible observable effects in astrophysics as well as in the laboratory.
In the same context also graviton-photon conversion on spin 0 and 1/2
particles was considered \cite{ravsun}.

To our knowledge, the photon-graviton amplitude has so far been
considered only at tree level. In the present paper, we extend its study
to include the one-loop corrections due to virtual spin 0 or 1/2  particles. 
Contrary to the tree level case, at one loop these amplitudes depend
nontrivially on both the photon energy and the background field strength.  
Thus it is conceivable that, for some
region in this two-parameter space, the one-loop amplitudes might be comparable
with or dominating over the tree level ones. 
In the first part of this article, we use the `string-inspired' worldline formalism to
obtain compact integral representations for these amplitudes. Their numerical study
will be undertaken in the second part.

Our calculation provides also a first example for the application of the string-inspired
technique to mixed photon-graviton amplitudes. In recent years, methods derived from 
\cite{berkos} or
inspired by \cite{strassler1} string theory have been extensively used for the calculation of
on-shell gluon \cite{bediko5glu} and graviton \cite{bedush,dunnor} amplitudes, as well as for QED amplitudes in
a constant field \cite{ss1,cadhdu,adlsch,rss,gussho,vv,jhep, marusc}. 
More recently, in \cite{baszir} 
also the contribution to the graviton vacuum polarization involving a scalar loop was obtained. 
The treatment of gravitational backgrounds in the wordline formalism involves a number of
mathematical subtleties which have been clarified only recently 
\cite{Bastianelli:1992be,Bastianelli:1993ct,DeBoer:1995hv,Bastianelli:1998jm,kleche,Bastianelli:2000nm,Bastianelli:2002qw}.

Phenomenologically, the photon-graviton amplitude is primarily of interest in the magnetic field case.
However, in the formalism used here it makes technically no essential difference
whether one calculates an amplitude in a constant magnetic or in a general
constant field. Thus we will keep the electric component.

The organization of the paper is simple: In chapter \ref{formalism}
we present the general formalism
for calculating one loop amplitudes involving either a scalar or spinor loop and any number
of photons and gravitons, in vacuum or in a constant external field.
Chapter \ref{scalarloop} contains the scalar loop calculation, chapter \ref{spinorloop}
the spinor loop one. Our conclusions are given in \ref{conclusions}. In the appendix
we verify that the results obey the relevant gauge and gravitational Ward identities.

\vspace{20pt}
\section{Mixed electromagnetic - gravitational amplitudes in 
the worldline formalism} 
\label{formalism}
\renewcommand{\theequation}{2.\arabic{equation}}
\setcounter{equation}{0}

The application of the worldline formalism to flat space calculations has been
described in detail in the review \cite{review}.
However, the generalization to processes involving curved backgrounds, or gravitons,
is less familiar. Therefore, after a brief description of the worldline
formalism adapted 
to the case of a constant electromagnetic background, we will describe   
certain subtleties arising in this formalism with the coupling to gravity.

Let us first consider the case of a scalar particle coupled to 
electromagnetism and gravity.  
We use here euclidean conventions. The QFT for this scalar
particle is described by a complex scalar field $\phi$ with action
\bear
S[\phi,\phi^* ;g,A] = \int d^Dx \sqrt{g}
\Big [g^{\mu\nu} (\partial_\mu -i e A_\mu) \phi^* (\partial_\nu + ie A_\nu)
\phi + (m^2 +\xi R) \phi^* \phi \Big ] 
\non\\
\label{s2.1}
\ear
where $\xi$ parameterizes an additional non-minimal coupling to the scalar 
curvature\footnote{ The value $\xi=0$ is the minimal coupling, 
while the value  $\xi= {D-2\over 4(D-1)}$ gives a 
conformally invariant coupling in the massless case.}. 
The corresponding one-loop effective action is formally given by
\footnote{Our present definition of the effective action differs by a sign
from the conventions in \cite{review}; however, there is no difference
at the amplitude level.}

\bear
e^{-\Gamma[g,A]} &=& \int {\cal D}\phi {\cal D}\phi^*
\ e^{-S[\phi,\phi^* ;g,A]}\nonumber\\
\Gamma[g,A] &=& -\log {\rm Det}^{-1} (-\square_A +\xi R+m^2) 
={\rm Tr} \log  (-\square_A +\xi R+m^2 )\non\\
\label{s2.2}
\ear
where $\square_A$ is the gauge and gravitationally covariant Laplacian for 
scalar fields obtained from (\ref{s2.1}). 
This effective action can be represented in the 
worldline formalism by
\bear
\Gamma[g,A]
= - \int_0^\infty {dT\over T } \int_{PBC} {\cal D}x\ e^{-S[x;g,A]}
\label{s2.3}
\ear
with the worldline action
\bear
S[x;g,A] = \int_{0}^{T} d\tau \Big (
{1\over 4 } g_{\mu\nu}(x) \dot x^\mu \dot x^\nu +
i e A_\mu(x) \dot x^\mu + \xi R(x) +m^2 \Big ) \ .
\label{s2.4}
\ear
This worldline representation contains a standard integral over the proper 
time $T$, and a quantum mechanical path integral over the particle coordinates
$x^\mu(\tau)$. Due to the trace in (\ref{s2.2}) 
the path integral is to be taken with periodic boundary conditions
$x^\mu(0)=x^\mu(T)$. Thus it corresponds to an integral over closed loops
in spacetime.
In the worldline or `string-inspired' formalism, the path integral (\ref{s2.3}) is
manipulated into Gaussian form and then evaluated using appropriate 
worldline correlators. 
In an amplitude calculation, the external fields are specialized to plane waves.
Then each external leg is represented by a vertex operator. For the electromagnetic
field this is the photon vertex operator

 \bear
V^A_{\rm scal}[k,\pol] &=&
\pol_{\alpha}\int_0^Td\tau\, \dot x^{\alpha}(\tau)\e^{ik\cdot x(\tau)}
\label{defVAscal}
\ear
with an associated coupling constant $-ie$.
The graviton vertex operator will be derived below.
For the calculation of the path integral,
one first splits off the loop average position:

\bear
x^{\mu}(\tau)&=&x_0^{\mu}+y^{\mu}(\tau)\, , \label{x0split}\\
x_0^{\mu} &\equiv&{1\over T}\int_0^Td\tau\, x^{\mu}(\tau).
\label{defx0}
\ear
The path integral then factors into
$\int Dx(\tau) = \int d^Dx_0\int Dy(\tau)$. The integral over
$x_0$ just produces the global delta function for
energy-momentum conservation. 
The reduced path integral $\int Dy(\tau)$ is Gaussian and can be calculated
using the Wick contraction rule \cite{strassler1}

\bear
\langle y^{\mu}(\tau_1)y^{\nu}(\tau_2)\rangle
 &=& -\delta^{\mu\nu}\,G_B(\tau_1,\tau_2).
\label{corrbos}
\ear
Here $G_B$ is the `worldline Green's function'

\bear
G_B(\tau_1,\tau_2)&=&\mid \tau_1-\tau_2\mid 
-{{(\tau_1-\tau_2)}^2\over T}, \non\\
\dot G_B(\tau_1,\tau_2) &=& {\rm sign}(\tau_1 - \tau_2)
- 2 {{(\tau_1 - \tau_2)}\over T},\nonumber\\
\ddot G_B(\tau_1,\tau_2)
&=& 2 {\delta}(\tau_1 - \tau_2)
- {2\over T}.\quad \nonumber\\
\label{GGdGdd}
\ear
Here and in the following we will often abbreviate 
$G_{B12}\equiv G_B(\tau_1,\tau_2)$ etc.,
and a `dot' generally refers to a derivative in the first variable.

The inclusion of a constant electromagnetic background
field $F_{\mu\nu}$ can be achieved easily \cite{ss1,rss}
using
Fock-Schwinger gauge centered at the loop average position $x_0$. 
In this gauge 

\bear
A_{\mu}(x)&=&{1\over 2} y^{\nu}F_{\nu\mu}
\label{FS}
\ear
so that the presence of the background field produces only
an additional term

\bear
\Delta L &=& {1\over 2}iey^{\mu}F_{\mu\nu}\dot y^{\nu}
\label{DeltaL}
\ear
to the worldline Lagrangian in (\ref{s2.4}).
Since this terms involves $y$ only quadratically it can be taken into
account by an appropriate change of the worldline correlators. Namely, instead
of (\ref{corrbos},\ref{GGdGdd}) one finds

\bear
\langle y^{\mu}(\tau_1)y^{\nu}(\tau_2)\rangle &=& - {\cal G}_B^{\mn}(\tau_1,\tau_2)
\label{corrbosF}
\ear
with a field dependent worldline correlator
\cite{shaisultanov,rss}

\bear
{\cal G}_{B}(\tau_1,\tau_2) &=&
{T\over 2{({\cal Z})}^2}\biggl({{\cal Z}\over{{\rm sin}({\cal Z})}}
{\rm e}^{-i{\cal Z}\dot G_{B12}}
+i{\cal Z}\dot G_{B12} -1\biggr),
\non\\
\dot{\cal G}_B(\tau_1,\tau_2)
&=&
{i\over {\cal Z}}\biggl({{\cal Z}\over{{\rm sin}({\cal Z})}}
{\rm e}^{-i{\cal Z}\dot G_{B12}}-1\biggr),
\nonumber\\
\ddot{\cal G}_{B}(\tau_1,\tau_2)
&=& 2\delta_{12} -{2\over T}{{\cal Z}\over{{\rm sin}({\cal Z})}}
{\rm e}^{-i{\cal Z}\dot G_{B12}}\,.\nonumber\\
\label{calGB}
\ear
where ${\cal Z}_{\mu\nu} \equiv eTF_{\mu\nu}$.
These expressions should be understood as power series in the
Lorentz matrix $\cal Z$; more explicit expressions are given below.
Note the symmetry properties

\bear
{\cal G}_{B12} &=& 
{\cal G}_{B21}^{T},
\quad
\dot{\cal G}_{B12} = 
-\dot{\cal G}_{B21}^{T},
\quad
{\ddot{\cal G}}_{B12} = 
{\ddot{\cal G}}_{B21}^{T}.
\non\\
\label{symmcalGBF}
\ear\no
Contrary to the vacuum case, in a constant field background the worldline
correlators have non-vanishing coincidence limits,

\bear
{\cal G}_{B}(\tau,\tau)&=&
{T\over 2{({\cal Z})}^2}
\biggl({\cal Z}\cot({\cal Z})-1
\biggr),
\non\\
\dot {\cal G}_B(\tau,\tau) &=& i{\rm cot}({\cal Z})
-{i\over {\cal Z}}\,.\non\\
\label{coincalG}
\end{eqnarray}
\noindent
In the following a `bar' on a quantity denotes
the subtraction of its coincidence limit, e.g.,

\bear
\overline {\cal G}_{B12} &\equiv& {\cal G}_{B12} - {\cal G}_{B11}\,.
\label{defbarGB}
\ear
Note that coincidence limits of worldline correlators are always constant due
to their translational invariance.

The path integral determinant also becomes field dependent. In flat space
\cite{ss1}

\bear
\int Dy(\tau)\,{\rm exp}
\biggl [- \int_0^T d\tau
\Bigl (\fourth {\dot x}^2 
+\half ie \,x^{\mu}F_{\mn}\dot x^{\nu} \Bigr)
\biggr]
\!&=&\!\!
\PITD {\rm det}^{-{1\over 2}}
\Bigl[{{\rm sin}({\cal Z})\over {\cal Z}}\Bigr].
\non\\
\label{freepiF}
\ear

The path integral representing
a spin $1\over 2$ particle in an electromagnetic field differs from  
the spin $0$ case (\ref{s2.3}) above by a global factor of $-{1\over 2}$, and
by an additional Grassmann path integral representing the spin,

\bear
\int_{ABC} {\cal D}\psi \,
{\rm exp}\biggl [- \int_0^T d\tau
\Bigl({1\over2}g_{\mn}(x)\psi^{\mu} \bigl(\dot \psi^{\nu}+\dot x^{\alpha}\Gamma^{\nu}_{\alpha\beta}(x)
\psi^{\beta}\bigr)  -i e \psi^\mu F_{\mu\nu}(x)\psi^\nu 
\Bigr)
\biggr ].
\non\\
\label{grassmannpi}
\ear
Here the path integral is over antiperiodic Grassmann functions, 
$\psi^{\mu}(T) = - \psi^{\mu}(0)$. In the vacuum case, the appropriate 
worldline correlator is

\bear
\langle \psi(\tau_1)\psi(\tau_2) \rangle &=&
{1\over 2}G_F(\tau_1,\tau_2) \equiv {1\over 2} {\rm sign}(\tau_1-\tau_2)\,.
\label{defGF}
\ear
For a constant background field this correlator turns into

\bear
\langle \psi^{\mu}(\tau_1)\psi^{\nu}(\tau_2)\rangle
&=& 
{1\over 2}{\cal G}_F^{\mu\nu}(\tau_1,\tau_2)\nonumber\\
\label{wickrulesf}
\ear
where

\bear
{\cal G}_{F}(\tau_1,\tau_2) &=& G_{F12}
{{\rm e}^{-i{\cal Z}\dot G_{B12}} \over {{\rm cos}({\cal Z})}},
\non\\
\dot{\cal G}_F(\tau_1,\tau_2)
&=& \dot G_{F12} + G_{F12} {2 i {\cal Z}\over{ T {\rm cos}({\cal Z})}}  
{\rm e}^{-i{\cal Z}\dot G_{B12}} \,. \non\\
\label{calGFGFd}
\ear
Its symmetry properties are

\bear
{\cal G}_F(\tau_1,\tau_2) = -{\cal G}_F^{T}(\tau_2,\tau_1),
\quad
\dot{\cal G}_F(\tau_1,\tau_2) =  
\dot{\cal G}_F^{T}(\tau_2,\tau_1) \ .
\ear
The coincidence limits are

\bear
{\cal G}_{F11} = -i \tan ({\cal Z})
\ ,  \ \ \ \  
\dot {\cal G}_{F11}  = 2\delta_{11} + 2 e F \tan ({\cal Z}) \, .
\ear
The free path integral in a constant $F_{\mu\nu}$ background in $D$
dimensions is normalized as

\bear
\int {\cal D}\psi \,
{\rm exp}\biggl [- \int_0^T d\tau
\Bigl({1\over2}\psi \cdot \dot \psi  -i e \psi^\mu F_{\mu\nu}\psi^\nu 
\Bigr)
\biggr ]
&=&
2^{D\over 2}
{\rm det}^{1\over 2}\Bigl[{\rm cos}({\cal Z})\Bigr].
\non\\
\label{grassmannpiF}
\ear
The photon vertex operator (\ref{defVAscal}) acquires an additional Grassmann piece,

\bear
V^A_{\rm spin}[k,\pol] &=&
\pol_{\alpha}\int_0^Td\tau \Bigl[ \dot x^{\alpha}(\tau)
+2 i \psi^\alpha(\tau) \psi(\tau)\cdot k \Bigr ]
\e^{ik\cdot x(\tau)}\,.
\non\\
\label{defVAspin}
\ear

In the flat space case, the naive Gaussian path integration gives unambiguous and
well-defined parameter integral representations. 
To the contrary, when the coupling to gravity is introduced more precise definitions 
of the worldline regularizations are required to define the path 
integral properly. Let us describe these issues briefly.
First of all it is convenient to exponentiate the nontrivial 
path integral measure in a regularization independent way 
by using ghost fields \cite{Bastianelli:1992be,Bastianelli:1993ct}.  
The covariant measure in (\ref{s2.3}) is of the form
\bear
{\cal D} x = Dx \prod_{ 0\leq \tau < T} \sqrt{\det g_{\mu\nu}(x(\tau))}   
\ear
where $Dx=\prod_\tau d^Dx(\tau)$ is the standard translationally 
invariant measure. It can be represented more conveniently by introducing
commuting $a^\mu$ and anticommuting $b^\mu , c^\mu$ ghosts with
periodic boundary conditions 
\bear
{\cal D} x = Dx \prod_{ 0 \leq \tau < 1} \sqrt{\det g_{\mu\nu}(x(\tau))}  =
Dx \int_{PBC} { D} a { D} b { D} c \;
{\rm e}^{- S_{gh}[x,a,b,c]} 
\ear
where the ghost action is given by
\bear
S_{gh}[x,a,b,c]
= \int_{0}^{T} d\tau \; {1\over 4}g_{\mu\nu}(x)(a^\mu a^\nu 
+ b^\mu c^\nu) \ . 
\ear
The extra vertices arising from the ghost action guarantee
that the final result will be finite (see, e.g.,  \cite{ly}).

For the perturbative calculation of graviton amplitudes
around flat space, one next linearizes the metric,

\bear
g_{\mu\nu}(x) &=& \delta_{\mu\nu} + \kappa h_{\mu\nu}(x)
\label{linearizeg}
\ear
and then specializes $h_{\mu\nu}(x)$ to plane wave form,

\bear
h_{\mu\nu}(x) &=& \varepsilon_{\mu\nu}\,\e^{ik\cdot x}\,.
\ear
This leads to the following vertex operator for the graviton coupled
to the loop scalar

\bear
V^h_{\rm scal}[k,\pol] &=& 
\pol_{\mn}\int_0^Td\tau \Bigl[\dot x^{\mu}(\tau)\dot x^{\nu}(\tau)
+a^{\mu}(\tau)a^{\nu}(\tau) + b^{\mu}(\tau)c^{\nu}(\tau)
\non\\&&\hspace{50pt}
+4\overline\xi (\delta^{\mn}k^2-k^{\mu}k^{\nu})\Bigr]
\,\e^{ik\cdot x(\tau)}
\non\\
\label{defVgravscal}
\ear
with an associated coupling constant factor of $-{\kappa\over 4}$.
The Wick contraction rules for the ghosts are

\bear
\langle a^{\mu}(\tau_1)a^{\nu}(\tau_2)\rangle
&=&
2\delta(\tau_1-\tau_2)\delta^{\mn},\non\\
\langle b^{\mu}(\tau_1)c^{\nu}(\tau_2)\rangle
&=&
-4 \delta(\tau_1-\tau_2)\delta^{\mn}.\non\\
\label{wickrulesghostscal}
\ear
Similarly, for the fermion loop case one finds a graviton
vertex operator

\bear
V^h_{\rm spin}[k,\pol] &=& 
\pol_{\mn}\int_0^Td\tau\Bigl[
\dot x^{\mu}(\tau)\dot x^{\nu}(\tau)
+a^{\mu}(\tau)a^{\nu}(\tau) + b^{\mu}(\tau)c^{\nu}(\tau)
\non\\&&\hskip -2cm 
+2\Big(\psi^\mu(\tau) \dot \psi^\nu(\tau) + \alpha^\mu(\tau) \alpha^\nu(\tau) 
+ i \dot x^{\mu}(\tau)\psi^\nu(\tau) \psi(\tau)\cdot k\Big)
\Bigr]
\,\e^{ik\cdot x(\tau)}
\non\\
\label{defVgravspin}
\ear
where $\alpha^\mu$ are the additional bosonic ghosts arising from
the nontrivial path integral measure for $\psi^\mu$ 
\cite{Bastianelli:2002qw}. 
Their correlator is

\bear
\langle \alpha^{\mu}(\tau_1)\alpha^{\nu}(\tau_2)\rangle
&=&
\delta(\tau_1-\tau_2)\delta^{\mn}.\non\\
\label{wickrulesghostspin}
\ear
In the perturbative expansion around flat space 
various worldline Feynman diagrams are linearly and logarithmically 
UV divergent.
In fact the additional derivative couplings due to the metric 
worsen power counting in momentum space.
The contributions from the measure, i.e. terms involving
the ghost correlators, will always eliminate 
these divergences.
Nevertheless, finite ambiguities are left over and dealt with 
by specifying a regularization scheme together with 
renormalization conditions 
\footnote{We emphasize that these issues concern the one-dimensional worldline theory
and bear no direct relation to the issue of regularization in spacetime,
to be considered later.} 
.
These amount to the requirement that the path integral in (\ref{s2.3})
correspond precisely to the Hamiltonian operator 
$H= (-\square_A +m^2 +\xi R)$ appearing in (\ref{s2.2}).
These renormalization conditions produce a finite counterterm 
of the form
\bear
\Delta S_{CT} = \int_{0}^{T} 
d\tau\, 2 \, V_{CT}
\ear
that must be added to the action (\ref{s2.4}).
Three regularization schemes have been worked out in detail
for this purpose: mode regularization (MR),
time slicing (TS), and 
dimensional regularization (DR).
The corresponding counterterms are
given by
\bear
V_{MR} &=& -{1\over 8} R + 
{1\over 8} g^{\mu\nu}\Gamma^\beta_{\mu\alpha} \Gamma^\alpha_{\nu\beta} \,,\non\\
V_{TS}&=& -{1\over 8} R 
-{1\over 24} g^{\mu\nu} g^{\alpha\beta} g_{\lambda\rho}
\Gamma^\lambda_{\mu\alpha} 
\Gamma^\rho_{\nu\beta} \,,\non\\
V_{DR} &=& -{1\over 8} R  \ . \non\\
\label{s2.9}
\ear
For details see \cite{Bastianelli:1998jm,DeBoer:1995hv,Bastianelli:2000nm}.

In the present article, we will be interested in the graviton-photon
amplitude in a constant electromagnetic background
mediated by a virtual scalar particle loop.
This means that we will only need to consider the linear coupling 
to the metric fluctuations around flat space. To this order the differences 
between the counterterms in (\ref{s2.9}) can be neglected 
since they are at least quadratic in metric fluctuations.
On the other hand the leading part of the counterterm contained in the
$-{1\over 8} R $ piece effectively changes the coupling
$\xi \to \bar \xi =\xi - {1\over 4}$ in (\ref{s2.4}).
This implies that to this order all three regularization schemes 
can be used interchangeably.

In the case of the spin $\half$ particle the coupling to $R$ is fixed by the Dirac
equation and corresponds to $\bar\xi =0$.
Additional details on the worldline formalism with background gravity 
can be found in \cite{baszir,Bastianelli:2002qw}.
 
\vspace{20pt}
\section{Calculation of the photon -- graviton amplitude
in a constant electromagnetic field: scalar loop}
\label{scalarloop}
\renewcommand{\theequation}{3.\arabic{equation}}
\setcounter{equation}{0}

According to the above, the amplitude for the interaction of
a graviton $h$ and a photon $A$ via a scalar loop, in the presence of
a constant electromagnetic background field $F_{\mn}$, is given by the
following expression:

\bear
\langle h(k_1)A(k_2)\rangle &=&
(-ie)(-{\kappa\over 4})
\Tintm
\non\\&&\times
\int_P 
Dx
DaDbDc
\,V^h_{\rm scal}[k_1,\pol^h] V^A_{\rm scal}[k_2,\pol^A]
\non\\&&\times
{\rm exp}\biggl [- \int_0^T d\tau
\Bigl (\fourth ({\dot x}^2 + a^2 + b\cdot c) 
+\half ie \,x^{\mu}F_{\mn}\dot x^{\nu} \Bigr)
\biggr ].
\non\\
\label{hAwl} 
\ear
Here $V^{A,h}_{\rm scal}$ represent the photon and graviton
vertex operators for the scalar loop case, (\ref{defVAscal}) and (\ref{defVgravscal}).

To start with, we perform the split (\ref{x0split}) and the trivial $x_0$ integration which
produces the global $\delta$ -- function for
momentum conservation.
The remaining path integral $\int Dy(\tau)$ is Gaussian and thus can be reduced
to Wick contraction. Taking (\ref{freepiF}) into account, one obtains

\bear
\langle h(k_1)A(k_2)\rangle &=&
{ie\kappa\over 4}
(2\pi)^D\delta(k_1+k_2)
\TintmD
{\rm det}^{-{1\over 2}}
\biggl[{{\rm sin}({\cal Z})\over {\cal Z}}\biggr]
\non\\&&\hspace{90pt}
\times
\Bigl\langle
V^h_{\rm scal}[k_1,\pol^h] V^A_{\rm scal}[k_2,\pol^A]
\Bigr\rangle
\non\\
\label{hAwlwick} 
\ear
(${\cal Z}_{\mn} = eTF_{\mn}$).

Performing the Wick contractions according to the rules
(\ref{corrbos}), (\ref{wickrulesghostscal}), and using
(\ref{symmcalGBF}), we obtain

\bear
\langle h(k_1)A(k_2)\rangle 
&=& (2\pi)^D\delta(k_1+k_2)
\pol^h_{\mn}\pol^A_{\alpha}\,\Pi^{\mu\nu ,\alpha}_{\rm scal}(k)
\label{defA}
\ear
where $k\equiv k_1$, and

\bear
\Pi^{\mu\nu ,\alpha}_{\rm scal}(k) &=&
{e\kappa\over 4 (4\pi)^{D\over 2}}
\Tintm T^{-{D\over 2}}
{\rm det}^{-{1\over 2}}
\Bigl[{{\rm sin}({\cal Z})\over {\cal Z}}\Bigr]
\non\\&&\times
\int_0^Td\tau_1\int_0^Td\tau_2
\,\e^{-k\cdot \overline {\cal G}_{B12}\cdot k}
I^{\mu\nu ,\alpha}_{\rm scal},
\non\\
\label{hAwlwickresultscal} 
\ear

\bear
I^{\mu\nu ,\alpha}_{\rm scal}
&=&
-\Bigl({\ddot {\cal G}}_{B11}^{\mn}-2\delta_{11}\delta^{\mn}\Bigr)
\Bigl(k\cdot \,{\overline {\dot {\cal G}}}_{B12}\Bigr)^{\alpha}
-\Bigl[ {\ddot {\cal G}}_{B12}^{\mu\alpha}
\Bigl(\,{\overline {\dot {\cal G}}}_{B12}\cdot k\Bigr)^{\nu}
+ \exmn \Bigr] \non\\&&
+ \Bigl(\,{\overline {\dot {\cal G}}}_{B12}\cdot k\Bigr)^{\mu}
\Bigl(\,{\overline {\dot {\cal G}}}_{B12}\cdot k\Bigr)^{\nu}
\Bigl(k\cdot \,{\overline {\dot {\cal G}}}_{B12}\Bigr)^{\alpha}
-4 {\bar\xi}(\delta^{\mn}k^2-k^{\mu}k^{\nu})
\Bigl(k\cdot \,{\overline {\dot {\cal G}}}_{B12}\Bigr)^{\alpha}.
\non\\
\label{Imunualpha}
\ear
It is useful to add to 
$I_{\rm scal}^{\mu\nu ,\alpha}\,\e^{-k\cdot \overline {\cal G}_{B12}\cdot k}$ 
the total derivative term

\bear
\half {\partial\over \partial\tau_1}
\Bigl[
\,{\overline {\dot {\cal G}}}_{B12}^{\mu\alpha}
\Bigl(\,{\overline {\dot {\cal G}}}_{B12}\cdot k\Bigr)^{\nu}
\,\e^{-k\cdot \overline {\cal G}_{B12}\cdot k}
+ \exmn
\Bigr].
\label{addtotder}
\ear
Then $I^{\mu\nu ,\alpha}_{\rm scal}$ 
gets replaced by $J^{\mu\nu ,\alpha}_{\rm scal}$,

\bear
J^{\mu\nu ,\alpha}_{\rm scal} &=&
J^{\mu\nu ,\alpha}_{{\rm scal}, 1} + J^{(\mu\nu) ,\alpha}_{{\rm scal}, 2} 
+ J^{(\mu\nu) ,\alpha}_{{\rm scal}, 3} 
+ J^{\mu\nu ,\alpha}_{{\rm scal}, 4}
\non\\
\label{decomposeJscal}
\ear
where $J^{(\mn)}=\half(J^{\mn}+J^{\nu\mu})$ and

\bear
J^{\mu\nu,\alpha}_{{\rm scal}, 1} &=&
-\Bigl({\ddot {\cal G}}_{B11}^{\mn}-2\delta_{11}\delta^{\mn}\Bigr)
\Bigl(k\cdot \,{\overline {\dot {\cal G}}}_{B12}\Bigr)^{\alpha}\, ,
\non\\
J^{\mu\nu,\alpha}_{{\rm scal}, 2} &=&
\,{\overline{\dot {\cal G}}}_{B12}^{\mu\alpha}
\Bigl( \ddot {\cal G}_{B12}\cdot k\Bigr)^{\nu}
-{\ddot {\cal G}}_{B12}^{\nu\alpha}
\Bigl(\,{\overline {\dot {\cal G}}}_{B12}\cdot k\Bigr)^{\mu}\, ,
\non\\
J^{\mu\nu,\alpha}_{{\rm scal}, 3} &=&
\Bigl(\,{\overline {\dot {\cal G}}}_{B12}\cdot k\Bigr)^{\mu}
\Bigl\lbrack
\Bigl(\,{\overline  {\dot {\cal G}}}_{B12}\cdot k\Bigr)^{\nu}
\Bigl(k\cdot \,{\overline {\dot {\cal G}}}_{B12}\Bigr)^{\alpha}
-
\,{\overline{\dot {\cal G}}}_{B12}^{\nu\alpha}
k\cdot {\dot {\cal G}}_{B12}\cdot k
\Bigr\rbrack \, ,
\non\\
J^{\mu\nu ,\alpha}_{{\rm scal}, 4} &=&
-4 {\bar\xi}(\delta^{\mn}k^2-k^{\mu}k^{\nu})
\Bigl(k\cdot \,{\overline {\dot {\cal G}}}_{B12}\Bigr)^{\alpha}.
\label{Jiscal}
\ear
Before proceeding further, let us use this integral representation
to analyse the general structure of this amplitude. Although
our calculation is nonperturbative in the external field,
we can, of course, use the series expansions 
of the worldline Green's functions (\ref{calGB}) to compute the amplitude
involving a given number of interactions with the field.
It is then immediately seen that this amplitude is nonzero only
if this number of interactions is odd, since otherwise the
$\tau_{1,2}$ integrations vanish by antisymmetry. 
Also, the integrand contains terms which are singular at $T=0$,
indicating UV divergences. While so far our calculation has been valid
for any spacetime dimension $D$, the structure of these divergences
depends on $D$. We therefore confine ourselves to the four dimensional
case in the following.
For $D=4 $ 
the terms $J_{{\rm scal}, 3,4}^{\mn,\alpha}$ are UV finite, while
$J_{{\rm scal}, 1,2}^{\mn,\alpha}$ contain terms with a logarithmic divergence
at $T=0$. Those divergent terms involve the field only linearly,
and are thus easy to compute by expanding the formulas (\ref{calGB})
to the linear order in $F$. 
In dimensional regularisation, the result is

\bear
\Pi_{\rm scal, div}^{\mn,\alpha}(k)
&=&
{ie^2\kappa\over 3(4\pi)^2}{1\over D-4}C^{\mn,\alpha}
\label{Adiv}
\ear
where

\bear
C^{\mn,\alpha} &=&
\bigl(F\cdot k\bigr)^{\alpha}\delta^{\mn}
+F^{\mu\alpha}k^{\nu}
+F^{\nu\alpha}k^{\mu}
-\bigl(F\cdot k\bigr)^{\mu}\delta^{\nu\alpha}
-\bigl(F\cdot k\bigr)^{\nu}\delta^{\mu\alpha} \, .
\non\\&&
\label{defCmna}
\ear
As expected, this counterterm is just the momentum space version
($f_\mn = k_{\mu}\varepsilon_{\nu} - k_{\nu}\varepsilon_{\mu}$)
of the tree level interaction term (\ref{treelevel}).

We perform renormalization by subtracting the amplitude
at zero field strength and zero momentum. This can be done under the
$T$ - integral, leading to the following form of the renormalized
amplitude $\bar \Pi$:

\bear
{\bar \Pi}^{\mu\nu ,\alpha}_{\rm scal}(k) &=&
{e\kappa\over 64 \pi^2}
\int_0^{\infty}{dT\over T^3}
\,\e^{-m^2T}\non\\&&\times
\biggl\lbrace
{\rm det}^{-{1\over 2}}
\Bigl[{{\rm sin}({\cal Z})\over {\cal Z}}\Bigr]
\int_0^Td\tau_1\int_0^Td\tau_2
\,\e^{-k\cdot \overline {\cal G}_{B12}\cdot k}
J^{\mu\nu ,\alpha}_{\rm scal}
+{2\over 3}iT^2 e C^{\mn,\alpha}
\biggr\rbrace
.
\non\\
\label{Ascalren} 
\ear
Further, again following \cite{vv,review} we split ${\cal G}_{Bij}$ into 

\bear
{\cal G}_B &=& {\cal S}_B + {\cal A}_B
\label{splitSA}
\ear
where ${\cal S}_B^{\mu\nu}$ contains the even powers of $F^{\mu\nu}$
in the power series representation of ${\cal G}_B^{\mu\nu}$, and
${\cal A}_B^{\mu\nu}$ the odd ones. After this replacement
all terms in the integrand are either symmetric or antisymmetric
under the exchange $\tau_1 \leftrightarrow \tau_2$, and the
antisymmetric ones can be deleted since their $\tau_{1,2}$ - integrals
vanish. Note that the exponent 
$k\cdot \overline {\cal G}_{B12}\cdot k=k\cdot\overline {\cal S}_{B12}\cdot k$ 
is symmetric under this exchange.

Further, as usual \cite{review} we rescale $\tau_i = Tu_i, i=1,2$, and use the
translation invariance in $\tau$ to set $u_2=0$.
After analytic
continuation to Minkowski spacetime,
\footnote{We use the metric tensor 
$(\eta^{\mn})={\rm diag}(-+++)$ in Minkowski spacetime.
The analytic continuation from Euclidean to Minkowski space amounts to substituting $\delta^{\mn}\to\eta^{\mn}$,
$k^4\to -ik^0$, $T\to is$.}
this leads to our final result for this amplitude:

\bear
{\bar \Pi}^{\mu\nu ,\alpha}_{\rm scal}(k) &=&
{e\kappa\over 64 \pi^2}
\int_0^{\infty}{ds\over s}
\,\e^{-ism^2}
\Biggl\lbrace
{\rm det}^{-{1\over 2}}
\Bigl[{{\rm sin}({\cal Z})\over {\cal Z}}\Bigr]
\int_{0}^1
du_1
\non\\&&\hspace{90pt}\times
\,\e^{-k\cdot {\overline{\cal S}_{B12}}\cdot k}
\sum_{m=1}^4
{\tilde J}^{(\mu\nu),\alpha}_{\rm scal,m}
+{2\over 3}ieC^{\mu\nu ,\alpha}
\Biggr\rbrace
\non\\
\label{Ascalfin} 
\ear
where
$C^{\mu\nu,\alpha}$ is as in (\ref{defCmna}) 
(with $\delta^{\mn}\to \eta^{\mn}$) and

\bear
\tilde J^{\mu\nu,\alpha}_{{\rm scal}, 1} &=&
-\Bigl({\ddot {\cal S}}_{B11}^{\mn}-2\delta_{11}\eta^{\mn}\Bigr)
\Bigl(k\cdot \,{\overline {\dot {\cal A}}}_{B12}\Bigr)^{\alpha}\, ,
\non\\
\tilde J^{\mu\nu,\alpha}_{{\rm scal}, 2} &=&
\,{\dot {\cal S}}_{B12}^{\mu\alpha}
\Bigl( \ddot {\cal A}_{B12}\cdot k\Bigr)^{\nu}
-{\ddot {\cal A}}_{B12}^{\nu\alpha}
\Bigl(\,{\dot {\cal S}}_{B12}\cdot k\Bigr)^{\mu}\, 
\non\\&&
+\,{\overline{\dot {\cal A}}}_{B12}^{\mu\alpha}
\Bigl( \ddot {\cal S}_{B12}\cdot k\Bigr)^{\nu}
-{\ddot {\cal S}}_{B12}^{\nu\alpha}
\Bigl(\,{\overline {\dot {\cal A}}}_{B12}\cdot k\Bigr)^{\mu}\, ,
\non\\
\tilde J^{\mu\nu,\alpha}_{{\rm scal}, 3} &=&
\Bigl(\,{\dot {\cal S}}_{B12}\cdot k\Bigr)^{\mu}
\Bigl\lbrack
\Bigl(\,{\dot {\cal S}}_{B12}\cdot k\Bigr)^{\nu}
\Bigl(k\cdot \,{\overline {\dot {\cal A}}}_{B12}\Bigr)^{\alpha}
+
\Bigl(\,{\overline  {\dot {\cal A}}}_{B12}\cdot k\Bigr)^{\nu}
\Bigl(k\cdot \,{\dot {\cal S}}_{B12}\Bigr)^{\alpha}
\non\\&&\hspace{65pt}
-
\,{\overline{\dot {\cal A}}}_{B12}^{\nu\alpha}
k\cdot {\dot {\cal S}}_{B12}\cdot k
\Bigr\rbrack
\non\\&&
+
\Bigl(\,{\overline {\dot {\cal A}}}_{B12}\cdot k\Bigr)^{\mu}
\Bigl\lbrack
\Bigl(\,{\dot {\cal S}}_{B12}\cdot k\Bigr)^{\nu}
\Bigl(k\cdot \,          {\dot {\cal S}}_{B12}\Bigr)^{\alpha}
+
\Bigl(\,{\overline  {\dot {\cal A}}}_{B12}\cdot k\Bigr)^{\nu}
\Bigl(k\cdot \,{\overline {\dot {\cal A}}}_{B12}\Bigr)^{\alpha}
\non\\&&\hspace{65pt}
-
\,{\dot {\cal S}}_{B12}^{\nu\alpha}
k\cdot {\dot {\cal S}}_{B12}\cdot k 
\Bigr\rbrack \, ,\non\\
\tilde J^{\mu\nu ,\alpha}_{{\rm scal}, 4} &=&
-4 {\bar\xi}(\eta^{\mn}k^2-k^{\mu}k^{\nu})
\Bigl(k\cdot \,{\overline {\dot {\cal A}}}_{B12}\Bigr)^{\alpha} \, .
\label{Jtildescal}
\ear
For applications of this amplitude it will be necessary to write
the matrix functions ${\cal S}$, ${\cal A}$ and their derivatives in more explicit
form. A suitable representation has been given in \cite{vv,review}. 
Let ${\cal F}=\half(B^2-E^2)$, ${\cal G}={\bf B}\cdot {\bf E}$ the two
Maxwell invariants and 

\bear
a \equiv \sqrt{\sqrt{{\cal F}^2+{\cal G}^2}+{\cal F}},
\qquad
b \equiv \sqrt{\sqrt{{\cal F}^2+{\cal G}^2}-{\cal F}}
\label{defab}
\ear
so that $a^2-b^2 = B^2-E^2$ and $ab={\bf E}\cdot{\bf B}$.
Let 

\bear
z_+ &\equiv& iesa, \qquad z_- \equiv -esb.
\label{defzpm}
\ear
Then the determinant factor decomposes as

\bear
{\rm det}^{-{1\over 2}}
\Bigl[{{\rm sin}({\cal Z})\over {\cal Z}}\Bigr]
&=&
{z_+z_-\over{\rm sinh}(z_+){\rm sinh}(z_-)} \, .
\label{decdet}
\ear
Let further

\bear
\hat{\cal Z}_+ &\equiv& {aF -b\tilde F\over a^2 + b^2} ,\qquad
\hat{\cal Z}_- \equiv -i{bF+a\tilde F\over a^2+b^2}, 
\label{defZ}
\ear
with $\tilde F_{\mu\nu} = \half \varepsilon_{\mu\nu\alpha\beta}F^{\alpha\beta}$
the dual field strength tensor\footnote{In our conventions $\varepsilon^{0123}=1$.}.
The matrices $\hat{\cal Z}_{\pm}$ fulfill

\bear
\hat{\cal Z}_+\cdot \hat{\cal Z}_- = 0
\label{Zorth}
\ear
and

\bear
\hat{\cal Z}_+^2 = {F^2-b^2\Eins \over a^2+b^2},\qquad
\hat{\cal Z}_-^2 = -{F^2+a^2\Eins\over a^2+b^2}.
\label{Zsquare}
\ear
Then one has the following orthogonal matrix decompositions:

\bear
{\cal S}_{B12} &=& i{s\over 2}\sum_{a=\pm}
{A^{a}_{B12}\over z_{a}}\hat{\cal Z}^2_{a},\nonumber\\
\dot{\cal S}_{B12} &=& -\sum_{a=\pm}S^{a}_{B12}\hat{\cal Z}^2_{a}
,\nonumber\\
\dot{\cal A}_{B12} &=& -i\sum_{a=\pm}A^{a}_{B12}\hat{\cal Z}_{a},
\nonumber\\
\ddot{\cal S}_{B12} &=&
\ddot G_{B12}\Eins -{2i\over s}\sum_{a=\pm}z_{a}A^{a}_{B12}
\hat{\cal Z}^2_{a},\nonumber\\
\ddot{\cal A}_{B12}&=&
{2\over s}\sum_{a=\pm}z_{a}S^{a}_{B12}\hat{\cal Z}_{a}.
\nonumber\\
\label{defcalSetc}
\ear
These formulas are written in terms of the following four basic scalar, dimensionless
coefficient functions:

\bear
S_{B12}^{\pm} &=&
{\sinh(z_{\pm}\dot G_{B12})\over \sinh(z_{\pm})} ,
\non\\
A_{B12}^{\pm} &=&
{\cosh(z_{\pm} \dot G_{B12})\over 
\sinh(z_{\pm})}-{1\over z_{\pm}} . \nonumber\\
\label{defSAB}
\ear
Note that with respect to the exchange $u_1\leftrightarrow u_2$ the $S^{\pm}_{B12}$ are
odd and the $A^{\pm}_{B12}$ even. Thus only
the latter have non-vanishing coincidence limits,

\bear
A_{Bii}^{\pm} &=&
\coth(z_{\pm})-{1\over z_{\pm}}\, .
\label{coinAB}
\ear

\vspace{20pt}
\section{Calculation of the photon -- graviton amplitude
in a constant electromagnetic field: spinor loop}
\label{spinorloop}
\renewcommand{\theequation}{4.\arabic{equation}}
\setcounter{equation}{0}

The corresponding calculation for the spinor loop case proceeds in a completely
analogous way:

\bear
\langle h(k_1)A(k_2)\rangle &=&
-{1\over 2}(-ie)(-{\kappa\over 4})
\Tintm
\int_P Dx DaDbDc
\int_A D\psi D\alpha 
\non\\&&
\hskip -3cm \times 
V^h_{\rm spin}[k_1,\pol^h] V^A_{\rm spin}[k_2,\pol^A]
\non\\&&
\hskip -3cm \times 
{\rm exp}\biggl [- \int_0^T d\tau
\Bigl (\fourth ({\dot x}^2 + a^2 + b\cdot c) 
+\half ie \,x^{\mu}F_{\mn}\dot x^{\nu} 
+{1\over2}(\psi \cdot \dot \psi +\alpha^2) -i e \psi^\mu F_{\mu\nu}\psi^\nu 
\Bigr)
\biggr ]
\non\\
&=&-{ie\kappa\over 8}
2^{D\over 2}
(2\pi)^D\delta(k_1+k_2)
\TintmD 
{\rm det}^{-{1\over 2}}
\biggl[{{\rm tan}({\cal Z})\over {\cal Z}}\biggr]
\non\\&&\hspace{90pt}
\times
\Bigl\langle
V^h_{\rm spin}[k_1,\pol^h] V^A_{\rm spin}[k_2,\pol^A]
\Bigr\rangle .
\non\\
\label{hAwlwickf} 
\ear
Here $V^{A,h}_{\rm spin}$ now represent the photon and graviton
vertex operators for the fermion loop case, (\ref{defVAspin}) and (\ref{defVgravspin}).
The additional 
Wick contraction rules have been given in (\ref{wickrulesf}),(\ref{wickrulesghostspin}).
Performing the Wick contractions in (\ref{hAwlwickf}) 
we obtain the analogue of (\ref{hAwlwickresultscal}),

\bear
\Pi^{\mu\nu ,\alpha}_{\rm spin}(k) &=&
-{e\kappa 2^{D\over 2} \over 8 (4\pi)^{D\over 2}}
\Tintm T^{-{D\over 2}}
{\rm det}^{-{1\over 2}}
\Bigl[{{\rm tan}({\cal Z})\over {\cal Z}}\Bigr]
\non\\&&\times
\int_0^Td\tau_1\int_0^Td\tau_2
\,\e^{-k\cdot \overline {\cal G}_{B12}\cdot k}
I^{\mu\nu ,\alpha}_{\rm spin} \, .
\non\\
\label{hAwlwickresultspin} 
\ear
Here 

\bear
I^{\mu\nu ,\alpha}_{\rm spin} &=&I^{\mu\nu ,\alpha}_{\rm scal} (\bar\xi =0) 
+I_{\rm extra}^{(\mn),\alpha}
\label{decIspin}
\ear
where

\bear
I^{\mu\nu ,\alpha}_{\rm extra}
&=&
-\Bigl[{\dot {\cal G}}_{F11}^{\mn}-2\delta_{11}\delta^{\mn}
+
\Bigl(\,{ { {\cal G}}}_{F11}\cdot k\Bigr)^{\nu}
\Bigl(\,{\overline {\dot {\cal G}}}_{B12}\cdot k\Bigr)^{\mu}
\Bigr]
\Bigl(\,{ { {\cal G}}}_{F22}\cdot k\Bigr)^{\alpha}
\non\\&&
+
\Bigl(\,{ { {\cal G}}}_{F12}\cdot k\Bigr)^{\mu}
{ {\dot {\cal G}}}_{F12}^{\nu \alpha}
-
{ { {\cal G}}}_{F12}^{\mu\alpha}
\Bigl(\,{ {\dot {\cal G}}}_{F12}\cdot k\Bigr)^{\nu}
\non\\&&
+\Bigl[{{\cal G}}_{F12}^{\nu\alpha}
\Bigl(\,{ {k\cdot {\cal G}}}_{F12}\cdot k\Bigr)
-
\Bigl(\,{ { {\cal G}}}_{F12}\cdot k\Bigr)^{\nu}
\Bigl(\,{ {k\cdot {\cal G}}}_{F12}\Bigr)^\alpha \Bigr]
\Bigl(\,{\overline {\dot {\cal G}}}_{B12}\cdot k\Bigr)^{\mu}
\non\\&&
+
\Bigl({\dot {\cal G}}_{F11}^{\mn}-2\delta_{11}\delta^{\mn}\Bigr)
\Bigl(\, k\cdot {\overline {\dot {\cal G}}}_{B12}\Bigr)^{\alpha}
\non\\&&
+ \Bigl[
\ddot {\cal G}_{B11}^{\mu\nu}
-2\delta_{11}\delta^{\mn} -
\Bigl(\,{\overline {\dot {\cal G}}}_{B12}\cdot k\Bigr)^{\mu}
\Bigl(\,{\overline {\dot {\cal G}}}_{B12}\cdot k\Bigr)^{\nu}
\Bigr]
\Bigl(\,{ { {\cal G}}}_{F22}\cdot k\Bigr)^{\alpha} 
\non\\&&
-\Bigl [ 
{\ddot {\cal G}}_{B12}^{\mu\alpha} 
- \Bigl(\,{\overline {\dot {\cal G}}}_{B12}\cdot k\Bigr)^{\mu}
\Bigl(\, k\cdot {\overline {\dot {\cal G}}}_{B12}\Bigr)^{\alpha}
\Bigr ]
\Bigl(\,{ { {\cal G}}}_{F11}\cdot k\Bigr)^{\nu} \, .
\non\\
\label{Iextra}
\ear
It is useful to replace $I_{\rm spin}^{\mu\nu,\alpha}$ by

\bear 
J_{\rm spin}^{\mn,\alpha} \equiv J_{\rm scal}^{\mn,\alpha}(\bar\xi =0) + I^{(\mn),\alpha}_{\rm extra}
\, .
\label{defJspin}
\ear
 
The integrand can then be rearranged in the following way
(compare with (\ref{Jiscal}) for the scalar loop):

\bear
J_{\rm spin}^{\mn ,\alpha}
&=&
J_{{\rm spin} ,1}^{\mn ,\alpha}
+J_{{\rm spin} ,2}^{(\mn) ,\alpha}
+J_{{\rm spin} ,3}^{(\mn) ,\alpha}
\non\\
\label{decomposeIspin}
\ear
where

\bear
J_{{\rm spin} ,1}^{\mn ,\alpha}
&=&
\Bigl({\ddot {\cal G}}_{B11}^{\mn}-{\dot{\cal G}}_{F11}^{\mn}\Bigr)
\Bigl[
\bigl({\overline {\dot {\cal G}}}_{B21}+
{\cal G}_{F22}\bigr)\cdot k\Bigr]^{\alpha} \, ,
\non\\
J_{{\rm spin} ,2}^{\mn ,\alpha}
&=&
\,{\overline{\dot {\cal G}}}_{B12}^{\mu\alpha}
\Bigl( \ddot {\cal G}_{B12}\cdot k\Bigr)^{\nu}
-{\cal G}_{F12}^{\mu\alpha}
\Bigl(\dot {\cal G}_{F12}\cdot k\Bigr)^{\nu}
-{\ddot {\cal G}}_{B12}^{\nu\alpha}
\Bigl[\bigl({\overline {\dot {\cal G}}}_{B12}+{\cal G}_{F11}\bigr)\cdot k\Bigr]^{\mu} 
\non\\&&
+{\dot {\cal G}}_{F12}^{\nu\alpha}
\Bigl({\cal G}_{F12}\cdot k\Bigr)^{\mu} \, ,
\non\\
J_{{\rm spin} ,3}^{\mn ,\alpha}
&=&
-
\,\Bigl({\overline{\dot {\cal G}}}_{B12}\cdot k\Bigr)^{\mu}
\Bigl\lbrace
\Bigl[\bigl({\overline{\dot{\cal G}}}_{B12}+{\cal G}_{F11}\bigr)\cdot k\Bigr]^{\nu}
\Bigl[\bigl({\overline{\dot{\cal G}}}_{B21}+{\cal G}_{F22}\bigr)\cdot k\Bigr]^{\alpha}
\nonumber\\&&
-\Bigl({\cal G}_{F12}k\Bigr)^{\nu}
\Bigl({\cal G}_{F21}k\Bigr)^{\alpha}
+{\overline{\dot{\cal G}}}_{B12}^{\nu\alpha}k\cdot {\dot{\cal G}}_{B12}\cdot k 
-{\cal G}_{F12}^{\nu\alpha}k\cdot {\cal G}_{F12}\cdot k
\Bigr\rbrace.
\non\\
\label{Iispin}
\ear
As in the scalar loop case, renormalization requires only the subtraction of
a logarithmic divergence, yielding

\bear
{\bar \Pi}^{\mu\nu ,\alpha}_{\rm spin}(k) \!\!&=&\!\!
-{e\kappa \over  32 \pi^2}
\int_0^{\infty}{dT\over T^3}
\,\e^{-m^2T}\non\\&&
\hspace{-6mm}
\times \biggl\lbrace
{\rm det}^{-{1\over 2}}
\Bigl[{{\rm tan}({\cal Z})\over {\cal Z}}\Bigr]
\int_0^Td\tau_1\int_0^Td\tau_2
\,\e^{-k\cdot \overline {\cal G}_{B12}\cdot k}
J^{\mu\nu ,\alpha}_{\rm spin}
-{4\over 3}iT^2 e C^{\mn,\alpha}
\biggr\rbrace . 
\non\\
\ear
The tensor  $C^{\mn ,\alpha}$ appearing in the counterterm is the same as 
in the scalar loop case, eq. (\ref{defCmna}).

We now need also the analogue of the decomposition formulas (\ref{splitSA}), (\ref{defcalSetc}), 
(\ref{defSAB}) for ${\cal G}_F$,

\bear
{\cal G}_F &=& {\cal S}_F
+ {\cal A}_F,
\ear

\bear
{\cal S}_{F12} &=&-
\sum_{a =\pm}
S_{F12}^{a}\,{\hat{\cal Z}}_a^2 \, ,
\non\\
{\cal A}_{F12} &=& 
-i
\sum_{a =\pm}
A_{F12}^{a}\,{\hat{\cal Z}}_a \, ,
\non\\
\dot {\cal S}_{F12} &=&
\dot G_{F12}\,\Eins
- {2i\over s}  
\sum_{a =\pm} z_a
A_{F12}^{a}\,{\hat{\cal Z}}_a^2 \, ,
\non\\
\dot {\cal A}_{F12} &=& 
{2 \over s}
\sum_{a =\pm} z_a
S_{F12}^{a}\,{\hat{\cal Z}}_a \, ,
\ear
written in terms of the basic coefficient functions

\bear
S_{F12}^{\pm} &\equiv& G_{F12}
{\cosh(z_{\pm}\dot G_{B12})\over \cosh(z_{\pm})} \, ,
\non\\
A_{F12}^{\pm} &\equiv& G_{F12}
{\sinh(z_{\pm} \dot G_{B12})\over 
\cosh(z_{\pm})} \, ,\non\\
A_{F11}^{\pm} &=&
{\rm tanh}(z_{\pm})\, .\nonumber
\label{coinAF}
\ear
As in the bosonic case $S_{F12}^{\pm}$, ($A_{F12}^{\pm}$) are odd (even)
with respect to $1\leftrightarrow 2$.

Proceeding as in the scalar loop case, we obtain our final result for the amplitude:

\bear
{\bar \Pi}^{\mu\nu ,\alpha}_{\rm spin}(k) &=&
-{e\kappa\over 32 \pi^2}
\int_0^{\infty}{ds\over s}
\,\e^{-ism^2}
\Biggl\lbrace
{z_+z_-\over \tanh(z_+)\tanh(z_-)}
\int_{0}^1
du_1
\non\\&&\hspace{90pt}\times
\,\e^{-k\cdot {\overline{\cal S}_{B12}}\cdot k}
\sum_{m=1}^3
{\tilde J}^{(\mu\nu),\alpha}_{\rm spin,m}
-{4\over 3}ieC^{\mu\nu ,\alpha}
\Biggr\rbrace
\non\\
\label{Aspinfin} 
\ear
where

\bear
\tilde J^{\mu\nu,\alpha}_{{\rm spin}, 1} &=&
-\Bigl({\ddot {\cal S}}_{B11}^{\mn}-{\dot {\cal S}}_{F11}^{\mn})
\Bigl(k\cdot \,\bigl({\overline {\dot {\cal A}}}_{B12}+{\cal A}_{F22}\bigr)\Bigr)^{\alpha} \, ,
\non\\
\tilde J^{\mu\nu,\alpha}_{{\rm spin}, 2} &=&
\,{\dot {\cal S}}_{B12}^{\mu\alpha}
\Bigl( \ddot {\cal A}_{B12}\cdot k\Bigr)^{\nu}
-
{\cal S}_{F12}^{\mu\alpha}
\Bigl( \dot {\cal A}_{F12}\cdot k\Bigr)^{\nu}
\non\\&&
-{\ddot {\cal A}}_{B12}^{\nu\alpha}
\Bigl(\,{\dot {\cal S}}_{B12}\cdot k\Bigr)^{\mu}
+
{\dot {\cal A}}_{F12}^{\nu\alpha}
\Bigl( {\cal S}_{F12}\cdot k\Bigr)^{\mu}
\non\\&&
+\,{\overline{\dot {\cal A}}}_{B12}^{\mu\alpha}
\Bigl( \ddot {\cal S}_{B12}\cdot k\Bigr)^{\nu}
-
{\cal A}_{F12}^{\mu\alpha}
\Bigl( \dot {\cal S}_{F12}\cdot k\Bigr)^{\nu}
\non\\&&
-{\ddot {\cal S}}_{B12}^{\nu\alpha}
\Bigl(\,\bigl({\overline {\dot {\cal A}}}_{B12}+{\cal A}_{F11}\bigr)\cdot k\Bigr)^{\mu}
+
{\dot {\cal S}}_{F12}^{\nu\alpha}
\Bigl({\cal A}_{F12}\cdot k\Bigr)^{\mu} \, ,
\non\\
\tilde J^{\mu\nu,\alpha}_{{\rm spin}, 3} &=&
\Bigl(\,{\dot {\cal S}}_{B12}\cdot k\Bigr)^{\mu}
\Bigl\lbrack
\Bigl(\,{\dot {\cal S}}_{B12}\cdot k\Bigr)^{\nu}
\Bigl(k\cdot \, \bigl({\overline {\dot {\cal A}}}_{B12}+{\cal A}_{F11}\bigr)\Bigr)^{\alpha}
-
\Bigl(\,{\cal S}_{F12}\cdot k\Bigr)^{\nu}
\Bigl(k\cdot  {\cal A}_{F12} \Bigr)^{\alpha}
\non\\&&
+
\Bigl(\, \bigl({\overline  {\dot {\cal A}}}_{B12}+{\cal A}_{F11}\bigr)\cdot k\Bigr)^{\nu}
\Bigl(k\cdot \,{\dot {\cal S}}_{B12}\Bigr)^{\alpha}
-
\Bigl({\cal A}_{F12}\cdot k\Bigr)^{\nu}
\Bigl(k\cdot \,{\cal S}_{F12}\Bigr)^{\alpha}
\non\\&&\hspace{65pt}
-
\,{\overline{\dot {\cal A}}}_{B12}^{\nu\alpha}
k\cdot {\dot {\cal S}}_{B12}\cdot k
+
{\cal A}_{F12}^{\nu\alpha}
k\cdot {\cal S}_{F12}\cdot k
\Bigr\rbrack
\non\\&&
+
\Bigl(\,{\overline {\dot {\cal A}}}_{B12}\cdot k\Bigr)^{\mu}
\Bigl\lbrack
\Bigl(\,{\dot {\cal S}}_{B12}\cdot k\Bigr)^{\nu}
\Bigl(k\cdot \,          {\dot {\cal S}}_{B12}\Bigr)^{\alpha}
-
\Bigl({\cal S}_{F12}\cdot k\Bigr)^{\nu}
\Bigl(k\cdot \, {\cal S}_{F12}\Bigr)^{\alpha}
\non\\&&
+
\Bigl(\, \bigl({\overline  {\dot {\cal A}}}_{B12}+{\cal A}_{F11}\bigr)\cdot k\Bigr)^{\nu}
\Bigl(k\cdot \, \bigl({\overline {\dot {\cal A}}}_{B12}+{\cal A}_{F11}\bigr) \Bigr)^{\alpha}
-
\Bigl({\cal A}_{F12}\cdot k\Bigr)^{\nu}
\Bigl(k\cdot \, {\cal A}_{F12}\Bigr)^{\alpha}
\non\\&&\hspace{65pt}
-
\,{\dot {\cal S}}_{B12}^{\nu\alpha}
k\cdot {\dot {\cal S}}_{B12}\cdot k
+
{\cal S}_{F12}^{\nu\alpha}
k\cdot {\cal S}_{F12}\cdot k
\Bigr\rbrack \, .\non\\
\label{Jtildespin} 
\ear

\vspace{20pt}
\section{Conclusions}
\label{conclusions}
\renewcommand{\theequation}{5.\arabic{equation}}
\setcounter{equation}{0}
We have obtained compact explicit 
integral representations for the one-loop photon-graviton
amplitudes involving a scalar or spinor loop and a constant 
electromagnetic field.  The use of the constant field worldline
formalism along the lines of \cite{vv} has allowed us to achieve
this with modest calculational effort, and without
having to specialize to a special Lorentz frame. As usual in this
formalism the calculation
for the spinor loop case has been an extension of the scalar loop one.
Since the application of this formalism to gravitational backgrounds
is still novel we have presented the calculation in some detail.

We have verified that our results for these 
amplitudes obey the gravitational and gauge Ward identities.
As a further check, we have also shown that their low energy limits 
agree with the result of an effective action calculation \cite{giesha}.

In the upcoming second part of this paper we use our results for a
numerical study of the photon-graviton conversion process in a magnetic
field. 

\vspace{15pt}

\noindent
{\bf Acknowledgements:}
C.S. thanks H. Gies for helpful discussions.
Both authors thank Stefan Theisen and the Albert-Einstein Institute, Potsdam, for hospitality. 
C. S. also thanks
V. Villanueva and the Institute of Physics and Mathematics of UMSNH for hospitality.

\section{Appendix: Ward identities}
\label{appward}
\renewcommand{\theequation}{A.\arabic{equation}}
\setcounter{equation}{0}

In this appendix we derive the relevant Ward identities and verify that our 
graviton-photon amplitudes satisfy them. 
This provides a good check on the correctness of our calculations.

There are two types of Ward identities:
one that originates from gauge invariance and 
one that follows from reparametrization invariance
(general coordinate invariance).

Gauge transformations are defined by
\bear
\delta_G A_\mu = \partial_\mu \lambda \ , \quad
\delta_G g_{\mu\nu} = 0
\ear
with an arbitrary local parameter $\lambda $.
Then gauge invariance of the effective action
\bear
\delta_G \Gamma[g,A] =0
\ear
implies that 
\bear
\nabla_\mu \bigg (
{1\over \sqrt g }{\delta \Gamma\over \delta A_\mu } \bigg )=0  \ .
\ear
Similarly, infinitesimal reparametrizations are given by 
\bear
\delta_R A_\mu = \xi^\nu \partial_\nu A_\mu  + \partial_\mu \xi^\nu A_\nu
 \ , \quad \quad
\delta_R g_{\mu\nu} = \nabla_\mu \xi_\nu  + \nabla_\nu \xi_\mu
\ear
with arbitrary local parameters $\xi^\mu $. The 
invariance of the effective action 
\bear
\delta_R \Gamma[g,A] =0
\ear
now  implies
\bear
\nabla_\mu \bigg (
{2\over \sqrt g }{\delta \Gamma \over \delta g_{\mu\nu }}
+{1\over \sqrt g }{\delta \Gamma \over \delta A_{\mu}}
A^\nu \bigg )
- {1\over \sqrt g }
{\delta \Gamma \over \delta A_{\mu}} \nabla^\nu A_{\mu}
=0  \ .
\ear
The Ward identities thus obtained can be combined and written more 
conveniently using standard tensor calculus as follows
\bear
&& \partial_\mu  {\delta \Gamma\over \delta A_\mu } =0 \, ,
\label{gwi1}\\[2mm]
&& 2 \partial_\mu
{\delta \Gamma \over \delta g_{\mu\nu}}
+{\delta \Gamma \over \delta A_{\mu}} \partial_\mu  A^{\nu}
+ \Gamma_{\mu\lambda}^\nu \bigg ( 2
{\delta \Gamma \over \delta g_{\mu\lambda }}
+ { \delta \Gamma \over \delta A_\mu } A^\lambda \bigg )
- {\delta \Gamma \over \delta A_{\mu}} \nabla^\nu A_\mu=0 \ .
\nonumber
\ear

Now we are ready to consider the special case of the
graviton-photon correlation function in flat space and 
in a constant electromagnetic background  $F_{\mu\nu}$ described 
by the gauge potential $\bar A_\mu(x) = {1\over 2} x^ \nu F_{\nu\mu}$
\bear
\Gamma^{\mu\nu,\alpha}_{(x,y)}
\equiv  {\delta^2  \Gamma\over \delta g_{\mu\nu}(x) \delta A_\alpha(y)}
\Biggr |_{g_{\mu\nu} = \delta_{\mu\nu} , \,  A_\alpha = \bar A_\alpha} \ .
\label{Auno}
\ear
Taking functional derivatives on the general Ward identities
(\ref{gwi1}) to relate them to  eq. (\ref{Auno})
we obtain
\bear
&& \partial_\alpha^{(y)} \Gamma^{\mu\nu,\alpha}_{(x,y)}  = 0\, , 
\label{Adue} \\
&& 2 \partial_\mu^{(x)} \Gamma^{\mu\nu,\alpha}_{(x,y)}
+
{\delta^2  \Gamma\over \delta A_\mu(x) \delta A_\alpha(y)} \Biggr |
F_\mu{}^\nu  +
{\delta  \Gamma\over \delta A_\mu(x) } \Biggr |
(\delta^{\alpha\nu} \partial_\mu^{(x)} -
\delta^{\alpha}_\mu \partial^\nu_{(x)}) \delta^D(x-y)  = 0  \ .
\nonumber
 \ear
Now we Fourier transform these identities to momentum space 
\bear
\int dx_1 .. dx_n\, e^{ik_1x_1 +.. +ik_nx_n}\, \Gamma_{(x_1,.., x_n)} 
= (2\pi)^D \delta(k_1+..+k_n)\Gamma_{(k_1,.., k_n)} 
\nonumber
\ear
and from eqs. (\ref{Adue}) we obtain
\bear
&& k_\alpha \Gamma^{\mu\nu,\alpha}_{(k,-k)} = 0\, ,
\label{wi1} \\
&& 2 k_\mu \Gamma^{\mu\nu,\alpha}_{(k,-k)}
+ i \Gamma^{\mu,\alpha}_{(k,-k)} F_\mu{}^\nu
+ \Gamma^{\mu}_{(0)} (\delta^{\alpha\nu} k_\mu - \delta^{\alpha}_\mu k^\nu )
= 0 \ .
\nonumber
\ear
One may already notice that the term proportional to $\Gamma^{\mu}_{(0)}$ 
can be discarded, since 
it vanishes at zero momentum.
Thus we see that the gravitational Ward identity  
relates the graviton-photon amplitude to the photon-photon amplitude.
The latter has been 
calculated in the worldline formalism
in \cite{vv,review}.
The two point functions used above are simply related to the vacuum 
polarizations computed in the main text and in \cite{vv,review} as follows
\bear
\Gamma^{\mu,\alpha}_{(k,-k)} &=& 
-\Pi^{\mu,\alpha}(k)\, ,\cr
\kappa \Gamma^{\mu\nu,\alpha}_{(k,-k)}  &=&  -
\Pi^{\mn,\alpha}(k)\, . 
\ear
Thus the expected Ward identities are

\bear
k_{\alpha}\Pi^{\mn,\alpha}(k) &=& 0
\label{wardgauge}
\ear
and 

\bear
k_{\mu}\Pi^{\mn,\alpha}(k) &=& {i\over 2}\kappa F^\nu{}_\mu\Pi^{\mu\alpha}(k)\, .
\label{wardgrav}
\ear

Let us start with the amplitude due to a scalar loop.
These Ward identities are most easily checked using the form
(\ref{Jiscal}). To verify the gauge Ward identity 
(\ref{wardgauge}) note that $J_{{\rm scal} ,2,3,4}^{\mn,\alpha}$ 
vanish at the integrand level
when contracted with $k_{\alpha}$. This is not the case for
$k_{\alpha}J_{{\rm scal} ,1}^{\mn,\alpha}$, but for this term the integral
vanishes because 
$k\cdot \,{\overline {\dot {\cal G}}}_{B12}\cdot k$
is antisymmetric in $\tau_{1,2}$.

The verification of the gravitational Ward identity (\ref{wardgrav})
requires a bit more work, since here only 
$k_{\mu}J_{{\rm scal} ,4}^{\mn,\alpha}$
drops out at the integrand level. We can simplify it
by adding a suitable total derivative term:

\bear
k_{\mu}J^{\mu\nu ,\alpha}_{\rm scal}\,\e^{(\cdot)}
&\to &
k_{\mu}J^{\mu\nu ,\alpha}_{\rm scal}\,\e^{(\cdot)}
- \Big ( \ddot {\cal G}_{B11}^{\nu\alpha} -
2\delta_{11}\delta^{\nu\alpha} \Big )
{\partial\over\partial\tau_1} \,\e^{(\cdot )}
\non\\&&
+\half
{\partial\over\partial\tau_1}
\Biggl\lbrace
\biggl[
\Bigl(k\cdot \overline{\dot {\cal G}}_{B12}\Bigr)^{\alpha}
\Bigl(\overline{\dot{\cal G}}_{B12}\cdot k\Bigr)^{\nu}
-\overline{\dot{\cal G}}_{B12}^{\nu\alpha}
k\cdot \dot {\cal G}_{B12}\cdot k 
\biggr]
\,\e^{(\cdot )}
\Biggr\rbrace
\non\\
&&\hspace{-90pt}=
\biggl\lbrace
\Bigl(k\overline{\cdot \dot {\cal G}}_{B12}\Bigr)^{\alpha}
\Bigl(\bigl(\overline{\ddot{\cal G}}_{B12}+2\delta_{11}\Eins\bigr)
\cdot k\Bigr)^{\nu}
-\Bigl(\overline{\ddot{\cal G}}_{B12} + 2\delta_{11}\Eins\Bigr)^{\nu\alpha}
k\cdot \dot{\cal G}_{B12}\cdot k
\biggr\rbrace
\,\e^{(\cdot)}
\non\\
&&\hspace{-90pt}=
\biggl\lbrace
\Bigl(k\overline{\cdot \dot {\cal G}}_{B12}\Bigr)^{\alpha}
\Bigl(\bigl(\overline{\ddot{\cal G}_{B12}-\ddot G_{B12}}\bigr)
\cdot k\Bigr)^{\nu}
-\Bigl(\overline{\ddot{\cal G}_{B12}-\ddot G_{B12}}\Bigr)^{\nu\alpha}
k\cdot \dot{\cal G}_{B12}\cdot k
\biggr\rbrace
\,\e^{(\cdot)}\, .
\non\\
\label{addwardg}
\ear
Here in the last step we have used the fact that ${\overline{\dot{\cal G}}}_{B12}$
and $k\cdot \dot{\cal G}_{B12}\cdot k$ have vanishing coincident limits to replace
$2\delta_{11}$  by $\ddot G_{B11}-\ddot G_{B12} = 2\delta_{11}-2\delta_{12}$.

Next, let us write down the 
representation corresponding to (\ref{hAwlwickresultscal})
for the photon - photon amplitude in a constant field
\cite{vv,review}:

\bear
\Pi^{\mu\alpha}_{\rm scal}(k) &=&
- {e^2\over  (4\pi)^{D\over 2}}
\Tintm T^{-{D\over 2}}
{\rm det}^{-{1\over 2}}
\Bigl[{{\rm sin}({\cal Z})\over {\cal Z}}\Bigr]
\non\\&&\times
\int_0^Td\tau_1\int_0^Td\tau_2 \ 
\e^{-k\cdot \overline {\cal G}_{B12}\cdot k} \
I^{\mu\alpha}_{\rm scal} 
\non\\
\label{vvscal} 
\ear
 where
\bear
I^{\mu\alpha}_{\rm scal} =
\overline{\dot{\cal G}}_{B12}^{\mu\alpha}\,
k\cdot\dot{\cal G}_{B12}\cdot k
-
\Bigl(\overline{\dot{\cal G}}_{B12}\cdot k\Bigr)^{\mu}
\Bigl(k\cdot \overline{\dot{\cal G}}_{B12}\Bigr)^{\alpha} \ .
\label{Imascal}
\ear
Using (\ref{hAwlwickresultscal}), (\ref{addwardg}), and (\ref{vvscal}),
the gravitational Ward identity (\ref{wardgrav})
can now be easily verified using the matrix identity

\bear
\ddot{\cal G}_{B12} 
-\ddot G_{B12}
&=&
2ieF\cdot\dot{\cal G}_{B12}\, .
\label{idZGdot}
\ear

Let us now consider the amplitude due to a fermion loop.
The gauge Ward identity is again easily seen to be satisfied.
To check the more subtle gravitational Ward identities we need
the photon-photon amplitude in the constant electromagnetic background  
\cite{vv,review}

\bear
\Pi^{\mu\alpha}_{\rm spin}(k) &=& 
{e^2 2^{D\over 2} \over 2  (4\pi)^{D\over 2}}
\Tintm T^{-{D\over 2}}
{\rm det}^{-{1\over 2}}
\Bigl[{{\rm tan}({\cal Z})\over {\cal Z}}\Bigr]
\non\\&&\times
\int_0^Td\tau_1\int_0^Td\tau_2\
\e^{-k\cdot \overline {\cal G}_{B12}\cdot k}\
I^{\mu\alpha}_{\rm spin} 
\non\\
\label{vvspin} 
\ear
where 
\bear
I^{\mu\alpha}_{\rm spin} = I^{\mu\alpha}_{\rm scal} +I^{\mu\alpha}_{\rm extra}
\ear
with $ I^{\mu\alpha}_{\rm scal}$ given in (\ref{Imascal}) and 
\bear
I^{\mu\alpha}_{\rm extra} \!\! &=&\!\!
-  
{\cal G}_{F12}^{\mu\alpha}  
\Bigl(k\cdot {\cal G}_{F12}\cdot k \Bigr) \cr
&+&\!\!
\Bigl({\cal G}_{F12}\cdot k \Bigr)^\mu
\Bigl(k\cdot {\cal G}_{F12}\Bigr)^\alpha
+
\Bigl({\cal G}_{F11}\cdot k \Bigr)^\mu
\Bigl({\cal G}_{F22} \cdot k \Bigr)^\alpha \cr
&-&\!\!
\Bigl({\cal G}_{F11}\cdot k \Bigr)^\mu
\Bigl(k\cdot \overline{\dot{\cal G}}_{B12}\Bigr)^{\alpha}
+
\Bigl( \overline{\dot{\cal G}}_{B12} \cdot k \Bigr)^{\mu}
\Bigl({\cal G}_{F22}\cdot k \Bigr)^\alpha \, .
\ear

We essentially proved earlier that the scalar part satisfies the 
gravitational Ward identity. Thus we contract the remaining terms 
with $k^\mu$ and add a suitable total derivative 

\bear
k_\mu I^{(\mu\nu),\alpha}_{\rm extra}\, \e^{(\cdot)}
+ {\partial\over\partial\tau_1} T^{\nu\alpha}
\ear
with

\bear
T^{\nu\alpha} &=& {1\over 2} \biggl [  
\Bigl({\cal G}_{F11}\cdot k \Bigr)^\nu
\Bigl({\cal G}_{F22}\cdot k \Bigr)^\alpha
+ \Bigl({\cal G}_{F11}\cdot k \Bigr)^\nu
\Bigl(k\cdot \overline{\dot{\cal G}}_{B12}\Bigr)^{\alpha}
\non\\
&+&  
{\cal G}_{F12}^{\nu\alpha}  
\Bigl(k\cdot {\cal G}_{F12}\cdot k \Bigr) -
\Bigl({\cal G}_{F12}\cdot k \Bigr)^\nu
\Bigl(k\cdot {\cal G}_{F22} \Bigr)^\alpha
\non\\
&-& 2
\Bigl(\overline{\dot{\cal G}}_{B12}\cdot k \Bigr)^\nu
\Bigl({\cal G}_{F22}\cdot k \Bigr)^\alpha
\biggr ] \e^{(\cdot)}
\ear
to obtain

\bear
k_\mu I^{(\mu\nu),\alpha}_{\rm extra}\, \e^{(\cdot)}
+ {\partial\over\partial\tau_1} T^{\nu\alpha} &=&
-\biggl [
\Bigl(k\cdot \overline{\ddot{\cal G}}_{B12}\Bigr)^\nu  
+
\Bigl(k\cdot \dot{\cal G}_{F11}\Bigr)^\nu  
\biggr ] \Bigl({\cal G}_{F22}\cdot k \Bigr)^\alpha
\non\\&&
+
\dot {\cal G}_{F12}^{\nu\alpha}  
\Bigl(k\cdot {\cal G}_{F12}\cdot k \Bigr)  
-
\Bigl ( \dot {\cal G}_{F12} \cdot k \Bigr)^\nu  
\Bigl(k\cdot {\cal G}_{F12} \Bigr)^\alpha   
\non\\ &&
+
\biggl [
\Bigl(k\cdot \dot{\cal G}_{F11}\Bigr)^\nu  
- 2 \delta_{11} k^\nu
\biggr ]  
\Bigl(k\cdot \overline{\dot{\cal G}}_{B12}\Bigr)^{\alpha} \, .
\non\\
\ear
Now using  

\bear
\ddot {\cal G}_{B12} &=&  
\ddot G_{B12} + 2ie F \dot {\cal G}_{B12}  \, ,
\non\\
\dot {\cal G}_{F12} &=&  
2 \delta_{12} + 2ie F {\cal G}_{F12}  
\ear
one can show that eq. (\ref{wardgrav}) holds.

\vfill\eject

%%%%%%%%%%%%%%%%%%%%%%%%%%%%%%%%%%%%%%%%%%%%%%%%%%%%%%%%%%%%%%%%%%%%%%%%%%%%%%%

\end{document}